\documentclass[preprint,showpacs,preprintnumbers,amsmath,amssymb]{revtex4}

\usepackage{graphicx}% Include figure files
\usepackage{epsfig}
\usepackage{epsf}

\usepackage{dcolumn}% Align table columns on decimal point
\usepackage{bm}% bold math

\newcommand{\eq}{\begin{equation}}
\newcommand{\ee}{\end{equation}}
\newcommand{\ea}{\begin{eqnarray}}
\newcommand{\eea}{\end{eqnarray}}

\newcommand{\cC}{\mathcal C}
\newcommand{\cO}{\mathcal O}

\begin{document}

\title{Baryonic Flux in Quenched and Two-Flavor Dynamical QCD
After Abelian Projection}

\author{V.G.~Bornyakov,$^{\rm 1,2,3}$ H.~Ichie,$^{\rm 3,4}$ Y.~Mori,$^{\rm 3}$
D.~Pleiter,$^{\rm 5}$ M.I.~Polikarpov,$^{\rm 2}$
G.~Schierholz,$^{\rm 5,6}$ T.~Streuer,$^{\rm 5,7}$ H.~St\"uben$^{\rm 8}$ and 
T.~Suzuki$^{\rm 3}$ }
\affiliation{
$~^{1}$Institute for High Energy Physics, RU-142281 Protvino, Russia\\
$~^{2}$ITEP, B.Cheremushkinskaya 25, RU-117259 Moscow, Russia\\
$~^{3}$Institute for Theoretical Physics, Kanazawa University, Kanazawa 920-1192, Japan\\
$~^{4}$Tokyo Institute of Technology, Ohokayama 2-12-1, Tokyo 152-8551, Japan\\
$~^{5}$NIC/DESY Zeuthen, Platanenallee 6, D-15738 Zeuthen, Germany\\
$~^{6}$Deutsches Elektronen-Synchrotron DESY D-22603 Hamburg, Germany\\
$~^{7}$Institut f\"ur Theoretische Physik, Freie Universit\"at Berlin, 
D-14196 Berlin, Germany \\
$~^{8}$Konrad-Zuse-Zentrum f\"ur Informationstechnik Berlin, D-14195 Berlin, Germany
}

\author{DIK collaboration}

\date{\today}
%%%%%%%%%%%%%%%%%%%%%%%%%%%%%%%%%%%%%%%%%%%%%%%%%%%%%%%%%%%%%%%%%%%%%%%
\begin{abstract}
We study the distribution of color electric flux of the three-quark system in 
quenched and full QCD (with $N_f = 2$ flavors of dynamical quarks) at zero and
finite temperature. To reduce ultra-violet fluctuations, the calculations are
done in the abelian projected theory fixed to the maximally abelian gauge. In
the confined phase we find clear evidence for a {\sf Y}--shape flux tube
surrounded and formed by the solenoidal monopole current, in accordance with
the dual superconductor picture of confinement. In the deconfined, high
temperature phase monopoles cease to condense, and the distribution of the
color electric field becomes Coulomb--like.
\end{abstract}

\maketitle

\section{Introduction}

So far most investigations of the static potential, and the dynamics that
drives it, have concentrated on the quark-antiquark ($Q\bar{Q}$) system, while
little is known about the forces of the three-quark ($3Q$) ensemble. For 
understanding the structure of baryons and, in particular, for modelling the
nucleon~\cite{Isgur}, it is important to learn about the forces and the 
distribution of color electric flux in the $3Q$ system as well.
A particularly interesting question is whether a genuine three-body force
exists and the confining flux tube is of {\sf Y}--shape, or whether the 
long-range potential is simply the sum of two-body potentials, in agreement
with a $\Delta$--Ansatz, resulting in a flux tube of $\Delta$--shape.
By a flux tube of {\sf Y}-- and $\Delta$--shape we understand a 
flux tube between the three quarks having shortest possible length and
a junction, and a flux tube constructed out of three 
quark-antiquark flux tubes taken with a factor $\frac{1}{2}$.

Several lattice quenched QCD studies report evidence for a $\Delta$--type 
long-range potential~\cite{bali,aft}, while others claim a genuine three-body 
force~\cite{tmns,afj}. 
In Ref.~\cite{tmns} various patterns of the three-quark
system were considered with the distance between quarks
in an equilateral triangle, $d$, up to 0.8 fm.
It  was found that at large distances
the $Y$--Ansatz gives a better description of the three-quark
potential than the $\Delta$--Ansatz. 
On the other hand, the authors of Ref.~\cite{afj} found that at 
distances $d < 0.7$ fm the
three-quark potential is described quite well by $\Delta$--Ansatz,
while it rises like the  $Y$--Ansatz at larger distances, $0.7<d<1.5$ fm.

The $Y$--Ansatz is also being supported by the
field correlator method~\cite{simonov}. The difference between a $\Delta$-- and 
{\sf Y}--shape potential is rather small and difficult to detect, because the
underlying Wilson loop decays approximately exponentially with increasing 
interquark distance. A computation of the distribution of the color electric 
flux inside the baryon might help to resolve this problem.

In this paper we shall study the static potential and the flux tube of the
$3Q$ system. The long-distance physics appears to be predominantly abelian -- 
being the result of a yet unresolved mechanism -- and driven by monopole
condensation. The use of abelian variables is an essential ingredient in our
work, as it leads to a substantial reduction of the statistical noise.
Preliminary results of this investigation have been reported in
Ref.~\cite{ibss}.

The paper is organized as follows. In Section~2 we describe the details of our
simulation, including the correlation functions that we are going to
compute. The results of the calculation are presented in Sections 3 and
4. Section 3 is devoted to the study of the $3Q$ system at zero temperature,
while Section 4 deals with the finite temperature case. Finally, in Section 5
we conclude.

\section{Simulation details}

We employ the Wilson gauge field action throughout this paper. In our studies
of full QCD we are using non-perturbatively $O(a)$ improved Wilson fermions,
\begin{equation}
S_F = S_F^{(0)} - \frac{{\rm i}}{2}\kappa g\, c_{SW} a^5 \sum_s {\bar{\psi}(s)
\sigma_{\mu\nu} F_{\mu\nu}(s)\psi(s)}\,,
\end{equation}
with $N_f=2$ flavors of dynamical quarks, where $S_F^{(0)}$ is the ordinary 
Wilson fermion action. Further details of the dynamical runs are given 
in~\cite{Hinnerk,zeroT}.

\begin{figure}[tphb]
\vspace*{0.5cm}
\begin{center}
\epsfig{file=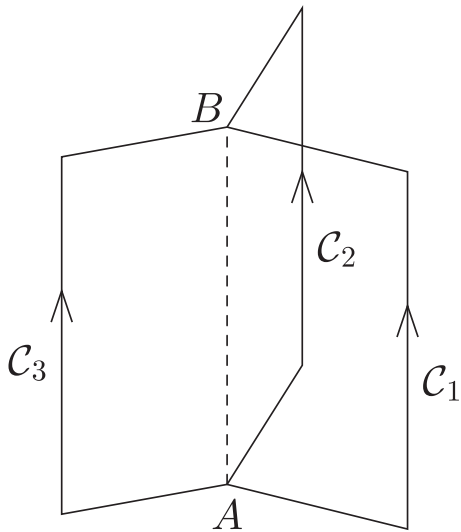,width=6cm,height=7cm,clip=} 
\vspace*{0.25cm}
\caption{Three quark Wilson loop.}
\end{center}
\vspace*{0.5cm}
\label{fig:Wil2}
\end{figure}

The system of three static quarks propagating from $A$ to $B$ may be described
by the `baryonic' Wilson loop
\begin{equation}
W_{3Q} = \frac{1}{3!}\, \varepsilon_{ijk}\varepsilon_{i'j'k'} U_{ii'}(\cC_1)
U_{jj'}(\cC_2) U_{kk'}(\cC_3)\,,
\end{equation}
where
\begin{equation}
U(\cC) = \prod_{s,\mu \in \cC} U(s,\mu)
\label{prod}
\end{equation}
is the ordered product of link 
matrices $U \in SU(3)$ along the path $\cC$, as shown in Fig.\ref{fig:Wil2}. 
The potential energy of this system is given by
\begin{equation}
V = - \frac{1}{L_T} \lim_{L_T \to \infty} \log\,\langle W_{3Q}\rangle \,,
\end{equation}
$L_T$ being the temporal extent of the loop.

%figure 1

In the following we shall concentrate on abelian variables, referring to the
maximally abelian gauge (MAG), and being obtained by standard abelian 
projection~\cite{tHooft,KLSW}. To 
fix the MAG~\cite{brand}, we use a simulated annealing algorithm described 
in~\cite{zeroT}. We write the abelian link variables as
\begin{equation}
\begin{split}
u(s,\mu) &\equiv {\rm diag}\big(u_1(s,\mu), u_2(s,\mu), u_3(s,\mu)\big)\,,\\[0.5em]
u_i(s,\mu)& = \exp({\rm i}\, \theta_i(s,\mu))
\end{split}
\end{equation}
with
\begin{equation}
\label{theta}
\begin{split}
 \theta_i(s,\mu) &= \arg(U_{ii}(s,\mu))-\frac{1}{3} \sum_{j=1}^3
\arg(U_{jj}(s,\mu))\,\big|_{\,{\rm mod}\ 2\pi}\,,  \\
&\theta_i(s,\mu) \in [-\frac{4}{3}\pi, \frac{4}{3}\pi]\,.
\end{split}
\end{equation}
They take values in $U(1)\times U(1)$, and under 
a general gauge transformation they transform as 
\begin{equation}
\label{trans}
\begin{split}
u(s,\mu) &\to d(s)^\dagger u(s,\mu) d(s+\hat{\mu})\,, \\
d(s) ={\rm diag}\big(
&\exp({\rm i}\,\alpha_1(s)),\exp({\rm i}\,\alpha_2(s)),\\
&\exp(-{\rm i}(\alpha_1(s)+\alpha_2(s)))\big)\,.
\end{split}
\end{equation}
The abelian Wilson loop is given by
\begin{equation} \label{W3Qab}
W_{3Q}^{\rm ab} = \frac{1}{3!}\, |\varepsilon_{ijk}| u_{i}(\cC_1)
u_{j}(\cC_2) u_{k}(\cC_3)\, ,
\end{equation}
where $u(\cC)$ is the abelian counterpart of (\ref{prod}). $W_{3Q}^{\rm ab}$ 
is invariant under gauge transformations (\ref{trans}).

The physical properties of the $3Q$ system can be infered from correlation 
functions of appropriate operators with the corresponding Wilson loop. 
Abelian operators take the form
\begin{equation}
\cO(s) =
\mbox{diag} (\cO_1(s),\cO_2(s),\cO_3(s)) \in U(1)\times U(1)\,.
\end{equation}
For C-parity even operators $\cO$, like the action and monopole densities, the
correlator of $\cO(s)$ with the abelian Wilson loop is given 
by~\cite{bikk,ibss}
\begin{equation}
\label{even}
\langle \cO (s) \rangle_{3Q} = \frac{\langle \cO(s) W_{3Q}^{\rm ab}
\rangle}{\langle W_{3Q}^{\rm ab} \rangle } - \langle \cO \rangle\,.
\end{equation}
For C-parity odd operators, like the electric field and monopole current which
carry a color index, the correlator is defined by
\begin{equation}
\label{odd}
\langle \cO (s) \rangle_{3Q} = \frac{1}{3!} \frac{\langle \cO_{i}(s)\,
|\varepsilon_{ijk}|\, u_{i}(\cC_1) u_{j}(\cC_2) u_{k}(\cC_3) \rangle} 
{\langle W_{3Q}^{\rm ab} \rangle}\, ,
\end{equation}
where summation over $i,j,k$ is assumed.
It is natural to use Wilson
loop to study the static potential at zero temperature since it gives
directly a singlet potential. The Polyakov loop correlator gives in
general a color-averaged potential, i.e. a mixture of the singlet and
octet potentials, see e.g. \cite{Nadkarni}. 
At nonzero temperature one can use only the Polyakov loop
correlator to study the static potential and we use the product
$P_{3Q}$ of three Polyakov loops closed around the boundary as baryonic
source instead of $W_{3Q}$:
\begin{equation} \label{P3q}
P_{3Q}^{\rm ab} = \frac{1}{3!}\, |\varepsilon_{ijk}|\, \ell_{i}(\vec{s}_1)\,
\ell_{j}(\vec{s}_2)\, \ell_{k}(\vec{s}_3)\, ,
\end{equation}
where
\begin{equation}
\ell_{i}(\vec{s}) = \prod_{s_4=1}^{L_T} u_{i}(\vec{s},s_4,4)
\end{equation}
is the abelian Polyakov loop, $L_T$ being the temporal extent of the lattice
here. The correlators of $\cO(s)$ with $P_{3Q}$ are defined analogously to 
(\ref{even}) and (\ref{odd}).

The observables we shall study are the action density $\rho_A^{3Q}$, 
the color electric field $E^{3Q}$ and the monopole current $k^{3Q}$.
The action density is given by
\begin{equation}
\rho_A^{3Q}(s) = \frac{\beta}{3} \sum_{i,\mu>\nu} \langle 
\cos(\theta_i(s,\mu,\nu))\rangle_{3Q} \,,
\end{equation}
where 
\begin{equation}
\begin{tabular}{c}
$\theta_i(s,\mu,\nu) = {\rm arg}(u_i(s,\mu,\nu))$\,, \\
$u_i(s,\mu,\nu) = u_i(s,\mu) u_i(s+\hat{\mu},\nu) 
u^\dagger_i(s+\hat{\nu},\mu) u^\dagger_i(s,\nu)$
\end{tabular}
\end{equation}
is the plaquette angle. The color electric field and monopole current
correlators are defined by
\begin{equation}
\label{electric}
E^{3Q}(s,i) = {\rm i}\, \langle \mbox{diag} (\theta_1(s,4,i),
\theta_2(s,4,i),\theta_3(s,4,i))\rangle_{3Q}
\end{equation}
and
\begin{equation}
k^{3Q}(^*s,\mu) = 2 \pi {\rm i}\, \langle \mbox{diag} 
(k_1(^*s,\mu),k_2(^*s,\mu),k_3(^*s,\mu)) \rangle_{3Q}\,, 
\end{equation}
where $k_i(^*s,\mu)$ is the monopole current~\cite{bikk,zeroT}.

The calculations in full QCD at zero temperature are performed on the 
$24^3\,48$ lattice at $\beta = 5.29$, $\kappa=0.1355$, which corresponds to 
a pion mass of $m_\pi/m_\rho \approx 0.7$ and a lattice spacing of $a/r_0 = 
0.18$~\cite{zeroT} (i.e. $a=0.09$ fm assuming $r_0=0.5$ fm). The calculations 
in full QCD at
finite temperature $T$ are done on the $16^3\,8$ lattice at $\beta=5.2$ for 
various hopping parameters ranging from $\kappa=0.1330$ to $\kappa=0.1360$, 
which covers the temperature range~\cite{finT} $0.8 \lesssim T/T_c \lesssim 
1.2$. The critical temperature $T_c$ corresponds to $\kappa=0.1344(1)$.
At this $\kappa$ we find $m_\pi/m_\rho \approx 0.77$.
For comparison, we also did quenched simulations at zero temperature on the 
$16^3\,32$ lattice at $\beta=6.0$. At this $\beta$ the lattice spacing
is $a/r_0=0.186$, i.e. it is roughly the same as on our full QCD lattices.
To reduce the statistical noise we smeared 
the spatial links of the abelian Wilson loop using 10 sweeps of APE
smearing \cite{Albanese:ds} with $\alpha=2$, where $\alpha$ is a 
coefficient multiplying the sum of staples.

\section{Static potential and baryonic flux at zero temperature}

The minimal {\sf Y}-type distance between the quarks,
i.e. the sum of the distance from the quarks to the Fermat point is \cite{tmns}
\begin{equation}
L_{\sf Y}=\left( \frac{1}{2}\sum_{i>j}\,r_{ij}^2+2\sqrt{3}\,S_{\Delta}
\right )
^{1/2},
\label{eq:R_min}
\end{equation}
where $\vec{r}_i$ marks the position of the $i^{\rm th}$ quark,
$r_{ij}=|\vec{r}_i-\vec{r}_j|$ and $S_{\Delta}$ is the area of the
triangle spanned by the three quarks.
The $Y$--Ansatz predicts that the confining part of the baryonic
potential is $\sigma^{3Q}_Y L_Y$, with string tension  $\sigma^{3Q}_Y$
equal to the $Q\bar{Q}$ string tension \cite{Carlson:1982xi}:
\begin{equation}
\sigma^{3Q}_Y = \sigma^{Q\bar{Q}}. 
\label{y-tension}
\end{equation}
The full expression describing both large and small distances is
\begin{equation}
V^{3Q}(L_Y)
 =  V_0^{3Q}   -  \sum_{i<j} \frac{\alpha^{3Q}}{r_{ij}}  + \sigma^{3Q}_Y L_Y \,,
\label{y-ansatz}
\end{equation}
where, similarly to the $Q\bar{Q}$ static potential, 
$V_0^{3Q}$ is a selfenergy term,  
the Coulomb term with effective coupling $\alpha^{3Q}$ comprises 
one gluon exchange as well as a L\"uscher term,
recently derived for the baryonic string in \cite{Jahn:2003uz},
and the confining term has string tension $\sigma^{3Q}_Y$.
The $\Delta$--Ansatz prediction \cite{Cornwall:1996xr} is that the 
confining part of the potential is proportional to the perimeter of the 
triangle formed by the quarks
\begin{equation}
L_{\Delta} = \sum_{i<j} | \vec{r}_i - \vec{r}_j|\,.
\end{equation}
with string tension 
\begin{equation}
\sigma^{3Q}_\Delta = \frac{1}{2}\sigma^{Q\bar{Q}}\,.
\label{d-tension}
\end{equation}
The short distance part is of the same form as in eq.(\ref{y-ansatz}).
Thus the full expression for the $\Delta$--Ansatz potential is
\begin{equation}
V^{3Q}(L_\Delta) =
V_0^{3Q} - \sum_{i<j} \frac{\alpha^{3Q}}{r_{ij}}  + \sigma^{3Q}_\Delta L_\Delta\,.
\label{d-ansatz}
\end{equation}
For short distances perturbation theory arguments 
relate the selfenergy and the Coulomb term coefficient
to those of the $Q\bar{Q}$ static potential \cite{afj}:
\begin{equation}
V_0^{3Q}=\frac{3}{2} V_0^{Q\bar{Q}}\,,\,\,\,\,\alpha^{3Q}=\frac{1}{2}
\alpha^{Q\bar{Q}}\,.
\label{32rule}
\end{equation}
On the other hand fitting the numerical data including both long
and short distances  by (\ref{y-ansatz}) or by (\ref{d-ansatz}) 
one may find results which differ from (\ref{32rule}), e.g. due
to the L\"uscher term contribution. In Ref.\cite{tmns} a rough agreement
between the fit parameters and (\ref{32rule}) has been found
for both $Y$--Ansatz and $\Delta$--Ansatz fits.

%figure 2
\begin{figure}[thpb]
\vspace*{0.5cm}
\begin{center}
\includegraphics[width=8.5cm,angle=0]{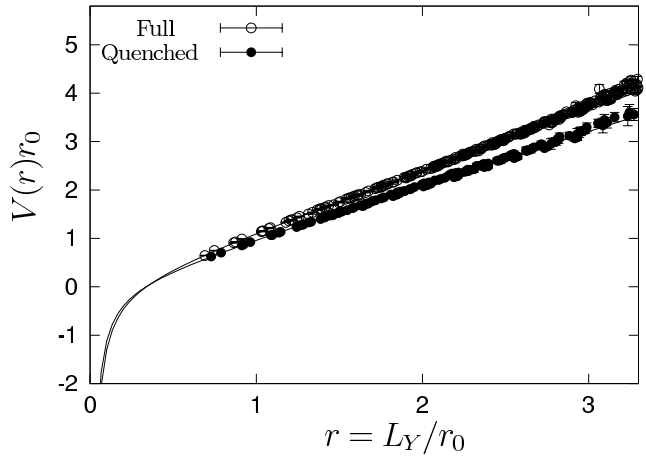} 
\caption{The abelian baryon potential in full and quenched QCD.} 
\label{fig:pot-q-f}
\end{center}
\vspace*{0.5cm}
\end{figure}
%figure 3
\begin{figure}[tphb]
\vspace*{0.5cm}
\begin{center}
\includegraphics[width=8.5cm,angle=0]{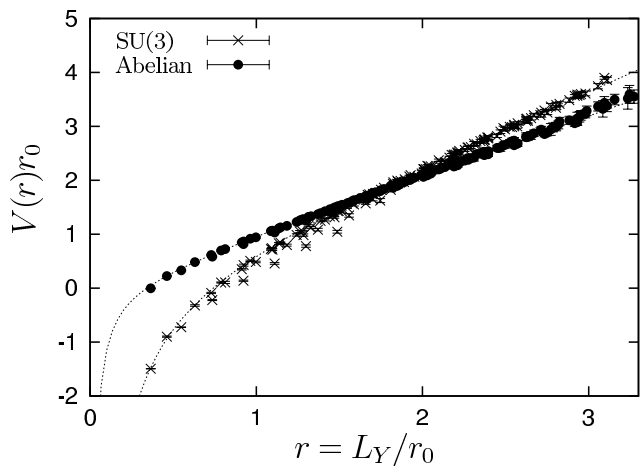} 
\caption{Comparison of the abelian and SU(3) baryon potential in the 
quenched approximation on the $16^3\,32$ lattice at $\beta=6.0$. The SU(3)
potential is taken from~\cite{tmns}.} 
\label{fig:potsu3ab}
\end{center}
\vspace*{0.5cm}
\end{figure}

In Fig.\ref{fig:pot-q-f} we show the baryon potential as a function of $L_Y$.
An unphysical constant $V^{3Q}_0$ has been subtracted from the 
potentials. For equal distances between the quarks, i.e. 
$|\vec{r}_i - \vec{r}_j| \equiv d = L_{\sf Y}/\sqrt{3}$ for $\forall\, 
i \neq j$, eq.(\ref{y-ansatz}) becomes
\begin{equation}
V^{3Q}(L_{\sf Y}) = V_0^{3Q} - 3 \sqrt{3} \frac{\alpha^{3Q}}{L_{\sf Y}} + 
\sigma^{3Q}_Y L_{\sf Y} \,.
\label{y-poten}
\end{equation}
Fitting our data for three quarks in equilateral triangles by 
eq.(\ref{y-poten}) for distances $d<0.75$ fm we found  the abelian string
tension $\sigma^{3Q}_{Y,\rm ab}\,a^2 = 0.038(1)$ and 0.0395(12) 
for the  quenched and full theory, respectively. These values agree
within error bars with the abelian string tension for the $Q\bar{Q}$
flux tube $\sigma_{\rm ab}^{Q\bar{Q}}=0.039(1)$ and 0.0402(11) \cite{zeroT}
thus supporting the $Y$--Ansatz. Note, that both $3Q$ and $Q\bar{Q}$
abelian string tensions are slightly higher in  full QCD.
We found values for the selfenergy and the Coulomb term
coefficient smaller than prescribed by (\ref{32rule}): 
%$V_0^{3Q} = 1.28(5) V_0^{Q\bar{Q}}\,,\, \alpha^{3Q} = 0.27(4)\alpha^{Q\bar{Q}}$
$V_0^{3Q}/V_0^{Q\bar{Q}} = 1.28(5),\, \alpha^{3Q}/\alpha^{Q\bar{Q}} = 0.27(4)$
in full QCD and 
%$V_0^{3Q} = 1.31(6) V_0^{Q\bar{Q}}\,,\, \alpha^{3Q} = 0.31(4)\alpha^{Q\bar{Q}}$
$V_0^{3Q}/V_0^{Q\bar{Q}} = 1.31(6),\, \alpha^{3Q}/\alpha^{Q\bar{Q}} = 0.31(4)$ 
in quenched QCD. 
The  fits are also shown in Fig.\ref{fig:pot-q-f}.

In Fig.\ref{fig:potsu3ab} the abelian and the nonabelian quenched potentials
are plotted together with respective fits. The data for the nonabelian
potential is taken from Ref.~\cite{tmns}. 
Comparison of $\sigma^{3Q}_{Y,\rm ab}$ with the SU(3) result~\cite{tmns} gives 
$\sigma^{3Q}_{Y,\rm ab}/\sigma^{3Q}_Y = 0.83(3)$, which lends 
further support to the hypothesis of abelian dominance.

%figure 4
\begin{figure}[tphb]
\vspace*{0.25cm}
\begin{center}
\includegraphics[width=8.5cm,angle=0]{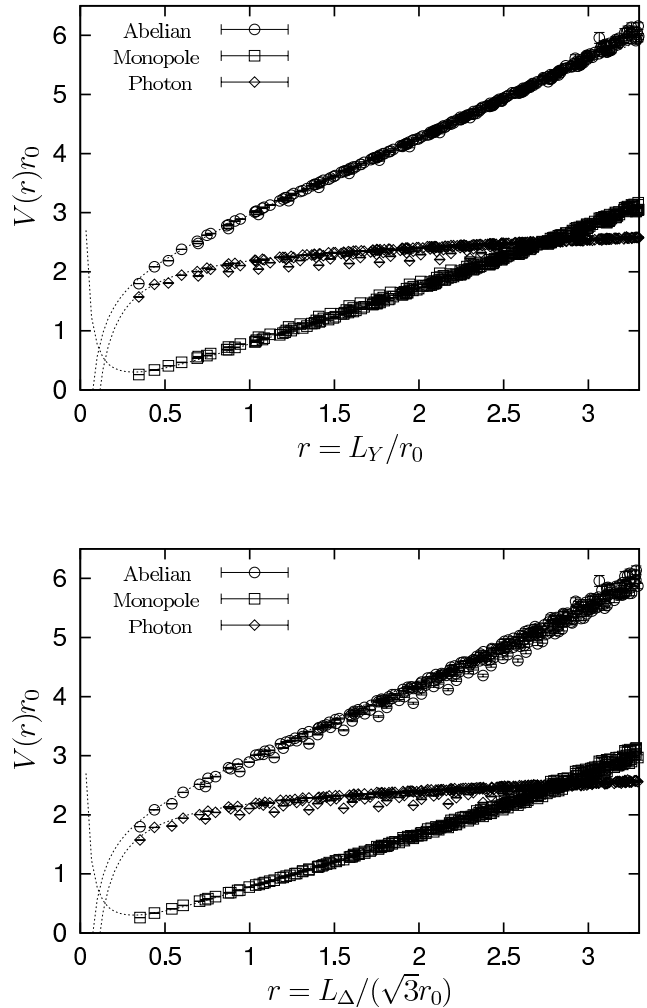}
\vspace*{0.25cm}
\caption{The abelian baryon potential in full QCD, together with its monopole 
and photon part, as a function of $L_{\sf Y}$ (top) and $L_\Delta$ (bottom),
respectively. The curves show fits (\ref{y-poten}) (top) and (\ref{d-ansatz})
(bottom) to the data with equal distances between the static sources.}
\label{fig:3qamp3}
\end{center}
\vspace*{-0.5cm}
\end{figure}

If the confining flux is of {\sf Y}-shape we would expect the long-distance
part of the potential to be a universal function of $L_{\sf Y}$. In 
Fig.\ref{fig:3qamp3} we plot the abelian potential as a function of 
$L_{\sf Y}$ (top), and as a function of $L_{\Delta}$ (bottom).
The data show a universal behavior when plotted against $L_{\sf Y}$. This is
to a lesser extent the case when plotted against $L_{\Delta}$, which supports
a genuine three-body force of {\sf Y}-type. 
In Fig.\ref{fig:3qamp3} the fits by the $ Y $--Ansatz and by
the $\Delta$--Ansatz for the quarks in the equilateral triangle are shown 
in the top and bottom parts, respectively.
Note that for the quarks in the equilateral triangle these two fits
are essentially the same with  
$\sigma^{3Q}_\Delta = \frac{1}{\sqrt{3}} \sigma^{3Q}_Y$ and equal selfenergy
and Coulomb coefficient.

%figure 5
\begin{figure}[htbp]
\begin{center}
\epsfig{file=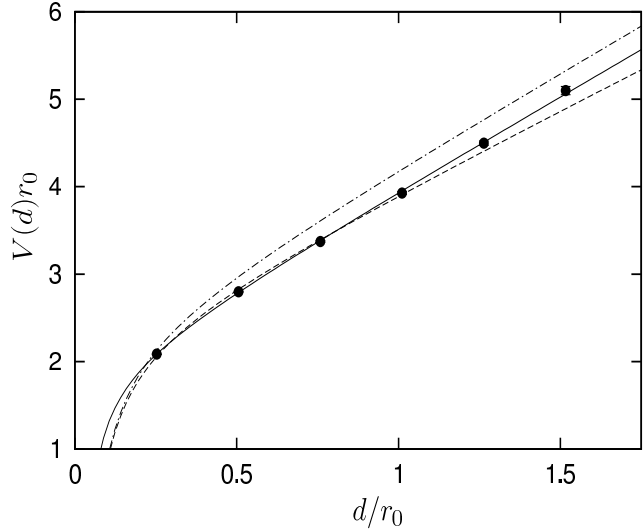,width=8.5cm,height=7cm}
%\vspace*{0.5cm}
\caption{The abelian three-quark potential and $\Delta$ and $Y$--Ans\"atze
in full QCD for quarks in equilateral triangles as function of quark 
separation $d=L_Y/\sqrt{3}=L_\Delta/3$. The solid line is a fit to the data, 
the dotted line is the
$\Delta$--Ansatz prediction eq. (\ref{d-ansatz}) and the dashed line is the 
$Y$--Ansatz prediction eq.(\ref{y-poten}). For both Ans\"atze  
eq.(\ref{32rule}) was used.}
\label{fig:comparison.poten}
\end{center}
\vspace*{0.25cm}
\end{figure}

%figure 6
\begin{figure}[htbp]
\begin{center}
\epsfig{file=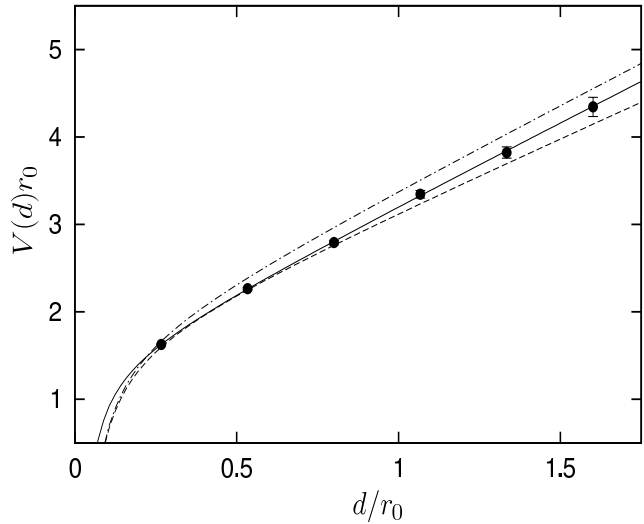,width=8.5cm,height=7cm}
\caption{Same as Fig.\ref{fig:comparison.poten} for quenched QCD.}
\label{fig:comparison.poten2}
\end{center}
\end{figure}
In Fig.\ref{fig:comparison.poten}  and Fig.\ref{fig:comparison.poten2} we show 
further comparison of our data with  $\Delta$ and $Y$--Ans\"atze
for full and quenched QCD.    
The data for the three-quark 
potential are plotted as a function of distance $d$ for equilateral triangle. 
In the same figures the curves showing respective $Y$--Ansatz and 
$\Delta$--Ansatz
predictions are plotted. We fix all parameters in the Ans\"atze using relations 
(\ref{y-tension}), (\ref{d-tension}) and  (\ref{32rule}).
We see that for both full and quenched QCD at distances $d < 0.5$ fm the three 
quark potential data agree with the $\Delta$--Ansatz, while at larger 
distances it agrees  with $Y$--Ansatz up to an additive constant
indicating that the string tension
$\sigma^{3Q}_{Y, \rm ab}$ is equal to $\sigma^{Q\bar{Q}}_{\rm ab}$ as was 
already discussed above. 
Similar findings were presented for the quenched nonabelian potential
in \cite{aft}. 
Thus we conclude that our data for the abelian potential confirm the
$Y$--Ansatz for large distances.
The agreement with the $\Delta$--Ansatz at short distances,
which was also observed in \cite{aft},  is probably a coincidence since  
the $\Delta$--Ansatz prediction eq.(\ref{d-tension}) is formulated for
large quark separations.  On the other hand, the 
proximity of the potential to the $\Delta$--Ansatz at distances which
are relevant for the spectrum calculations might be important for 
the phenomenologists since the calculations with the $\Delta$--Ansatz
potential are much simpler.
The disagreement with the Y-Anzatz at small distances was first clearly
observed in Ref.\cite{afj}.  One can guess that the finite size of the junction
play the role in appearance of this discrepancy. Although our data for the
static potential at small distances behave similar to that of Ref.\cite{afj} we
are not in a position to make strong statements about the short distances
since we are using the abelian projection which, as many earlier
observations suggest, gives correct description of the static potentials
at large distances only.
%figure 7
\begin{figure*}[tphb]
\vspace*{0.5cm}
\begin{center}
\epsfig{file=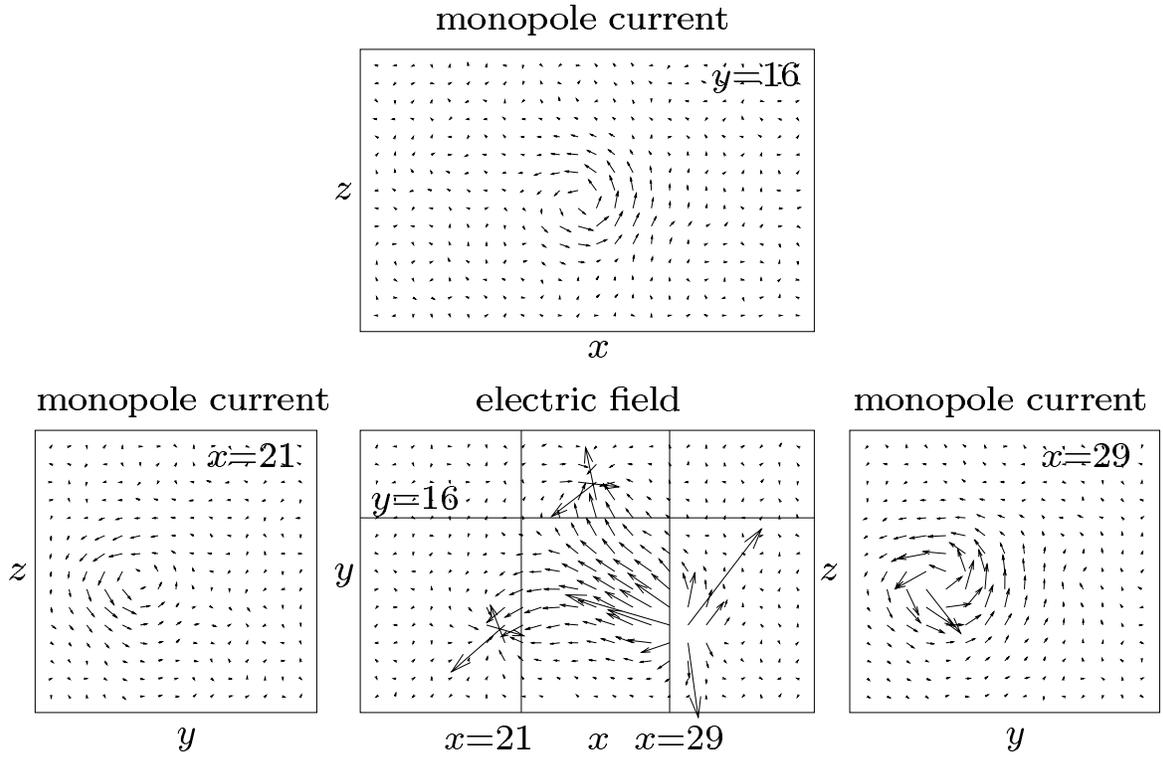,width=15.5cm,height=10cm,clip=}
\vspace*{0.5cm}
\caption{Distribution of the color electric field $\vec{E}^{3Q}$ in the 
$(x,y)$ plane on the $24^3\, 48$ lattice (center bottom figure), together 
with the monopole currents $k^{3Q}$ in the $(x,z)$ and $(y,z)$ 
planes (adjacent figures), respectively, at the position marked by the 
respective
solid lines. The magnitude of $E^{3Q}$ and $k^{3Q}$ is indicated by the length 
of the arrows.}
\label{fig:ERGB}
\end{center}
\vspace*{0.5cm}
\end{figure*}
%
%figure 8
\begin{figure}[tphb]
\begin{center}
\includegraphics[width=4cm,angle=0]{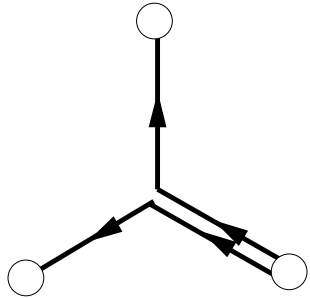} 
\vspace*{0.25cm}
\caption{Schematic view of the color electric field.}
\label{fig:ele}
\end{center}
\vspace*{-0.25cm}
\end{figure}

Although the results for the static potential are in favor of the 
$Y$--Ansatz, the difference from the $\Delta$--Ansatz prediction is rather
small. Thus it is worthwhile to study the color flux distribution.
In Fig.\ref{fig:ERGB} we show the distribution of the color electric 
field $\vec{E}^{3Q}$,
and its surrounding monopole currents $k^{3Q}$, on the $24^3\, 48$ lattice in 
full QCD. The time direction of the Wilson loop has been taken in one of
the spatial directions of the lattice. Points on the hyperplane orthogonal to 
the time direction of the Wilson loop are marked by $(x,y,z)$. The static 
quarks are placed at $(x,y,z)=(20,10,8)$, $(25,18,8)$ and $(30,10,8)$, 
respectively, i.e. they lie in the $(x,y)$ plane. The 
color index of the electric field operator (cf. eq.~(\ref{electric})) is 
identified with the color index of the quark in the bottom-right corner (in 
the center bottom figure). Note that the sum of the electric field over the 
three color indices vanishes at any point. As expected, the flux emanates from
the quark in the  
bottom-right corner and at about the center of the $3Q$ system splits into 
two parts. The flux lines are schematically drawn in Fig.\ref{fig:ele}.
A similar picture holds for the top and bottom-left quark and their respective
fluxes. 
In the adjacent figures we show the monopole 
current in the planes perpendicular to the electric flux lines, 
i.e. the $(x,z)$ and $(y,z)$ planes. They form a solenoidal current, as in 
the case of the $Q\bar{Q}$ system, in agreement with the dual superconductor 
picture of confinement.

We may decompose the abelian gauge field into a monopole and photon
part according to the definition~\cite{Smit,Suzuki}
\begin{equation}
\label{monph}
\begin{tabular}{c}
$\theta_i(s,\mu) = \theta_i^{\rm mon}(s,\mu)+\theta_i^{\rm ph}(s,\mu)$\,, \\
$\theta_i^{\rm mon}(s,\mu)= 2 \pi \sum_{s'} D(s-s')\nabla^{(-)}_{\alpha}
m_i(s',\alpha,\mu)$\,,
\end{tabular}
\end{equation}
where $D(s) = \Delta^{-1} (s)$ is the lattice Coulomb propagator,
$\nabla^{(-)}_{\mu}$ is the lattice backward derivative, and 
$m_i(s,\mu,\nu)$ counts the number of Dirac strings piercing the 
plaquette $u_i(s,\mu,\nu)$.
If one computes $k_i(^*s,\mu)$ from 
$\theta_i^{\rm mon}(s,\mu)$ one recovers almost all monopole currents.
In Fig.\ref{fig:3qamp3} we see that the monopole part is largely responsible 
for the linear behavior
of the potential, as was found already in case of the $Q\bar{Q}$ 
potential~\cite{zeroT}. The ratio of monopole to abelian string tension turns 
out to be 0.81(3).

\begin{figure}[thpb]
\begin{center}
\hspace{-4.0cm}
\epsfig{file=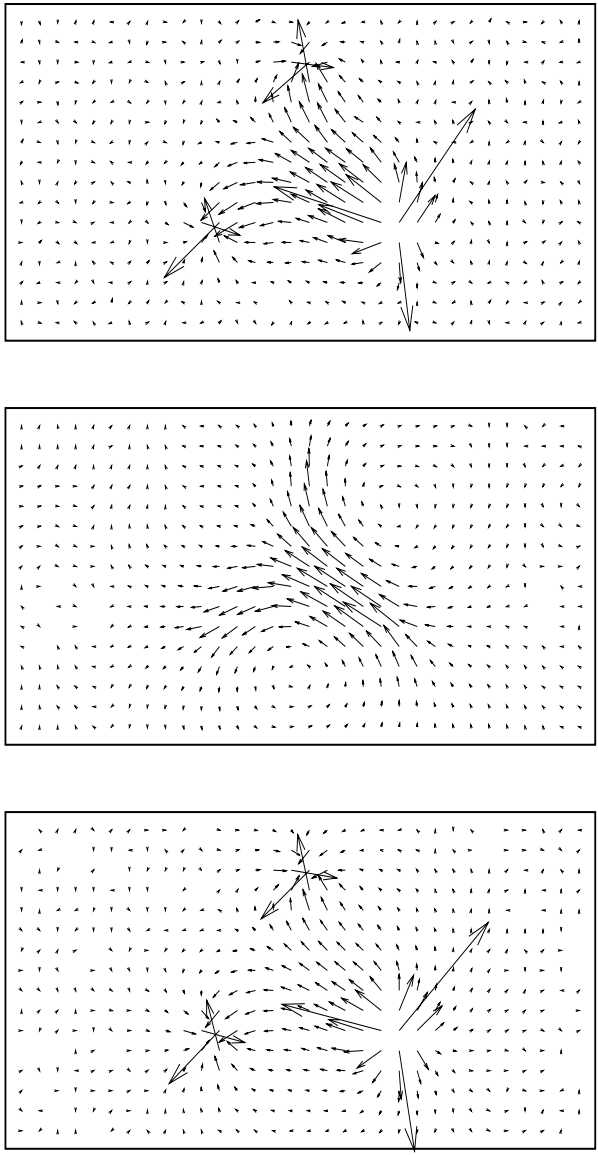,width=11.5cm,height=18.0cm,clip=} 
\vspace*{-2.0cm}
\caption{Distribution of the abelian color electric field $\vec{E}^{3Q}$
(top) broken into monopole (middle) and photon parts (bottom) on the 
$24^3\, 48$ lattice in full QCD. The color index of the electric field 
operators corresponds to that of the quark in the bottom right corner.
Only part of the lattice is shown here.}
\label{fig:EPMA}
\end{center}
\end{figure}

In Fig.\ref{fig:EPMA} we show the distribution of the abelian color electric 
field 
photon parts. The photon part shows a Coulomb-like distribution, while the
monopole part has no sources. Outside the flux tube the monopole and photon
parts of the color electric field largely cancel. The middle figure shows 
clearly that the flux lines are attracted to a {\sf Y}-type geometry.

\begin{figure*}[thpb]
\begin{center}
\epsfig{file=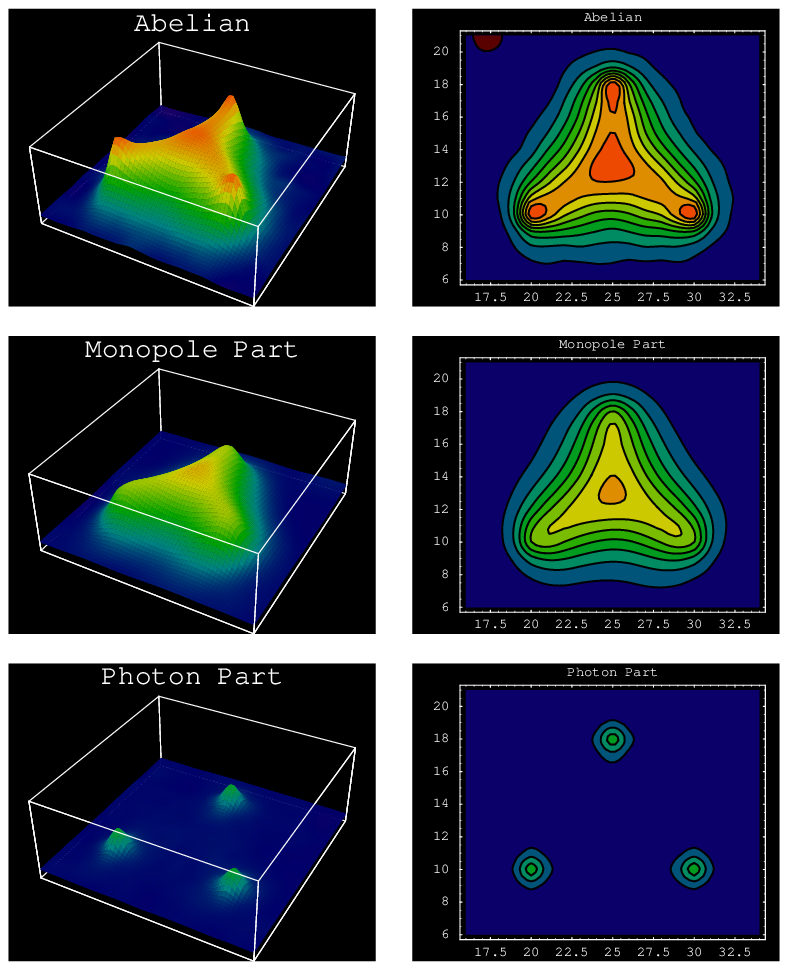,width=14.5cm,height=21cm}
\vspace*{0.5cm}
\caption{The abelian action density of the $3Q$ system in full QCD, together
with the monopole and photon part.}
\label{fig:PMA455action}
\end{center}
\vspace*{0.25cm}
\end{figure*}

In Fig.\ref{fig:PMA455action} we show the action density $\rho_A^{3Q}$ of 
the $3Q$ system in full QCD. Also shown is
the monopole and photon part of $\rho_A^{3Q}$ separately. Let us first look at
the (full)
abelian density. It clearly displays a {\sf Y}-type geometry of the color
forces. This is, of course, indistinguishable from
a geometry of purely two-body forces with strongly attracting flux lines. The
monopole part of the action density shows no sources. Apart from that, it 
appears that the 
action density originates almost entirely from the monopole part. The sources
show up in the photon part of the action density as expected.

%figure 9
%figure 10
\begin{figure}[htbp]
%\vspace*{-2.0cm}
\epsfig{file=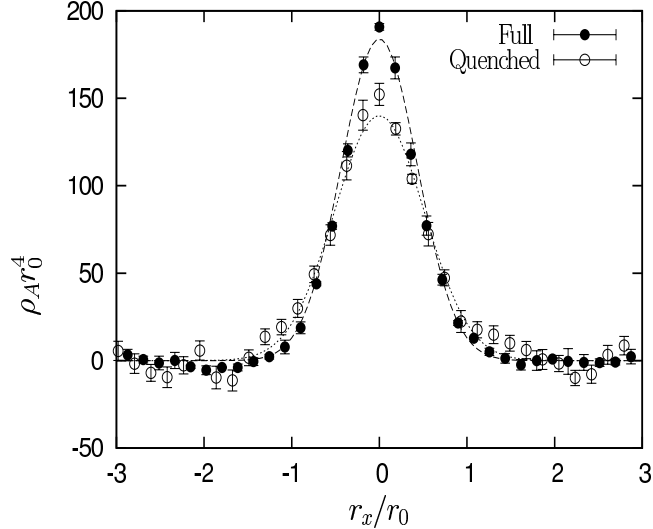,width=8.5cm,height=7.0cm,clip=} 
\begin{center}
\caption{The action density $\rho_A^{3Q}r_0^4$ of Fig.~\ref{fig:PMA455action} 
plotted across the flux tube at the distance $3a$ from the quark
and $2a$ from the junction.}
\label{fig:action.section}
\end{center}
\end{figure}

We have done similar calculations (to the ones shown in Figs.\ref{fig:3qamp3}, 
\ref{fig:ERGB}, \ref{fig:EPMA}, \ref{fig:PMA455action}) in 
quenched QCD as well. Part of our findings have been reported in~\cite{ibss}, 
and we refrain from repeating them here. Qualitatively, we found the same 
results as in full QCD at $m_\pi/m_\rho \approx 0.7$. 
In Fig.~\ref{fig:action.section} we compare the action
density of full and quenched QCD. 
We see that at the center of the flux tube the action density in full QCD is 
slightly higher than for the quenched case, while the shapes are rather similar.
The same feature has been observed for the $Q\bar{Q}$ flux tube \cite{zeroT}. 
We have estimated the width $\delta$ of the flux tube 
using a Gaussian fit \cite{zeroT}. The result is 
$\delta=0.30(4)$ fm and $0.36(11)$ fm in full and quenched QCD,
respectively. This is to be compared with the width of the
$Q\bar{Q}$ flux tube, which turned out to be
0.29(1) fm in full and quenched QCD \cite{zeroT}. We found that the 
width increases closer to the junction. So the numbers quoted
above are only to tell that the width of the baryon flux tube, away from the 
junction is not very different from that of the $Q\bar{Q}$ flux tube.
For a more precise determination of the width 
larger quark separation is necessary.

%figure 11

\begin{figure*}[tphb]
\begin{center}
\epsfig{file=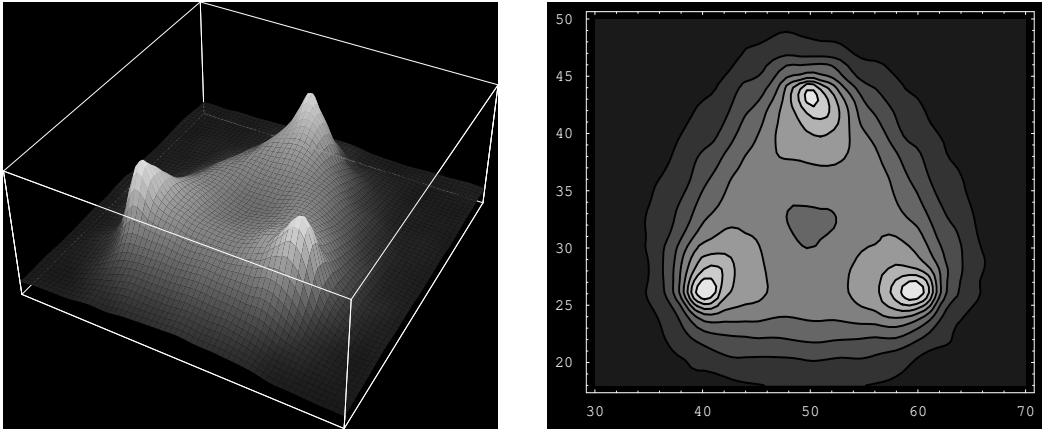,width=14.5cm,height=6cm}
\vspace*{0.5cm}
\caption{The abelian action density of the $3Q$ system as predicted by
the $\Delta$--Ansatz in full QCD. }
\label{fig:action.sum-q-aq}
\end{center}
\vspace*{0.25cm}
\end{figure*}

It is interesting to compare the action density shown in 
Fig.~\ref{fig:PMA455action}
with the action density constructed out of three $Q\bar{Q}$ flux tube       
action densities multiplied, in agreement with (\ref{d-tension}), by a factor 
$\frac{1}{2}$ to take into account that we are dealing with
pairs of quarks rather than with $Q\bar{Q}$ pairs. Such a comparison
has been done in Ref.\cite{afj} for the Potts model. For the $Q\bar{Q}$ 
action density we used the results of 
Ref.\cite{zeroT}. The resulting density
is shown in Fig.~\ref{fig:action.sum-q-aq}.  
Figures \ref{fig:PMA455action} and \ref{fig:action.sum-q-aq} look rather
different. The most important difference is that the measured density has a 
bump in the center, while the $\Delta$--Ansatz density has a dip. This 
comparison gives further support to the $Y$--Ansatz.

%figure 12

%\clearpage
\section{Baryonic flux at finite temperature}

We expect the flux tube to disappear and the color electric field to become 
Coulomb- or Yukawa-like above the finite temperature phase transition $T_c$
and when the string breaks in full QCD. This phenomenon has been observed in 
case of the $Q\bar{Q}$ system in the pure SU(2) gauge theory for temperatures
$T > T_c$~\cite{hay2} and in full QCD for $T$ just below and above 
$T_c$~\cite{bikm}. Throughout this Section we shall use the Polyakov loop 
(\ref{P3q}) to create a baryon.

\begin{figure}[tphb]
\begin{center}
\includegraphics[width=9.4cm,angle=0]{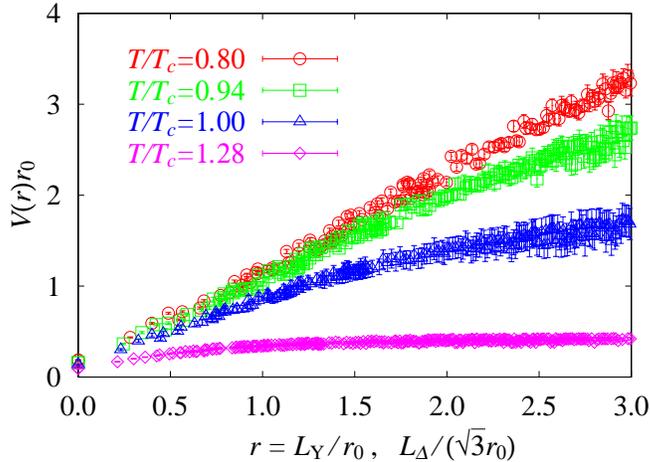} 
\vspace*{0.25cm}
\caption{The
monopole part of the baryon potential at finite temperature in full QCD as a 
function of $L_{\sf Y}$ ($T \le T_c$) and $L_{\Delta}$ ($T>T_c$), respectively.}
\label{fig:3qpotft}
\end{center}
\vspace*{0.5cm}
\end{figure}

In Fig.\ref{fig:3qpotft} we show the baryon potential on the $16^3\,8$ 
lattice at $\beta=5.2$ for several values of $\kappa$. At this $\beta$ value
\begin{equation}
T \propto \exp(-2.81/\kappa)\,.
\end{equation}
Increasing $\kappa$ thus increases the temperature. We cross the finite 
temperature phase transition at $\kappa = 0.1344$~\cite{finT}. We see that 
the potential flattens off while we approach the transition point. However, 
the distances we were able to probe are not large enough to make any statement
about string breaking. 

%figure 13

To compute the action density $\rho_A^{3Q}$ and the electric field
and monopole correlators $E^{3Q}$ and $k^{3Q}$, respectively, we need to
reduce the statistical noise. 
%Note that the Polyakov loops span an area of
%$\approx 16 \times 8$ lattice spacings. 
We do that by averaging over time slices and using extended operators
\begin{equation}
\begin{split}
\rho_A^{3Q}(s) &\longrightarrow  \\
&\frac{1}{8}
\{\rho_A^{3Q}(s) + \rho_A^{3Q}(s-\hat{ 1}-\hat{ 2}-\hat{ 3})
+ \rho_A^{3Q}(s-\hat{1}-\hat{ 2})\\ 
&+\rho_A^{3Q}\big(s-\hat{1}-\hat{3})
+ \rho_A^{3Q}(s-\hat{2}-\hat{3})
+ \rho_A^{3Q}(s-\hat{1})\\ 
&+\rho_A^{3Q}(s-\hat{2})
+ \rho_A^{3Q}(s-\hat{3})\}\,,
\end{split}
\end{equation}           
\begin{equation}
E^{3Q}(s,i)  \longrightarrow \frac{1}{2}\{E^{3Q}(s,i) + E^{3Q}(s-\hat{i},i) \}\,,
\end{equation}
\begin{equation}
k^{3Q}(^*s,\mu) \longrightarrow \frac{1}{2}\{k^{3Q}(^*s,\mu) +
k^{3Q}(^*(s-\hat{\rm 2}),\mu)\}\,,\,\mu=1,3
\end{equation}                         
where (again) we have assumed that the quarks lie in the $(x,y)$ plane, 
and we consider the monopole current in the $(z,x)$ plane.
%and we call the direction of the Polyakov lines the $t$ direction. 

\begin{figure*}[tphb]
\vspace{0.25cm}
\begin{center}
\epsfig{file=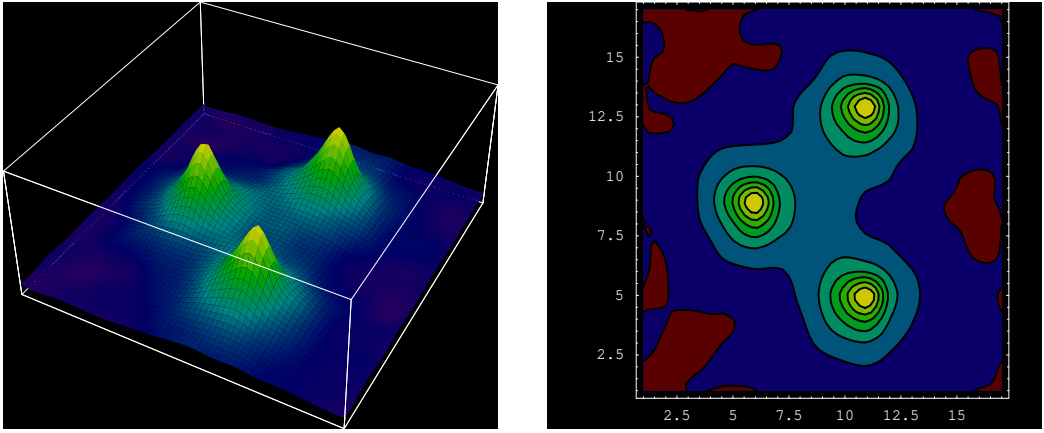,width=14.5cm,height=6cm}
\vspace*{0.5cm}
\caption{The Abelian action density in the
deconfined phase at $\kappa=0.1360$.} 
\label{fig:FTaction}
\end{center}
\end{figure*}

In Fig.\ref{fig:FTaction} we plot the abelian action density in the deconfined 
phase at $\kappa=0.1360$. As was to be expected, the action density shows three 
Coulomb-like peaks at the position of the quarks, similar to the photon part 
of the action density at zero temperature as shown in 
Fig.\ref{fig:PMA455action}.

%figure 14

%figure 15
\begin{figure*}[tphb]
\begin{center}
\epsfysize=9.5cm
\epsfxsize=14.5cm
\epsfbox{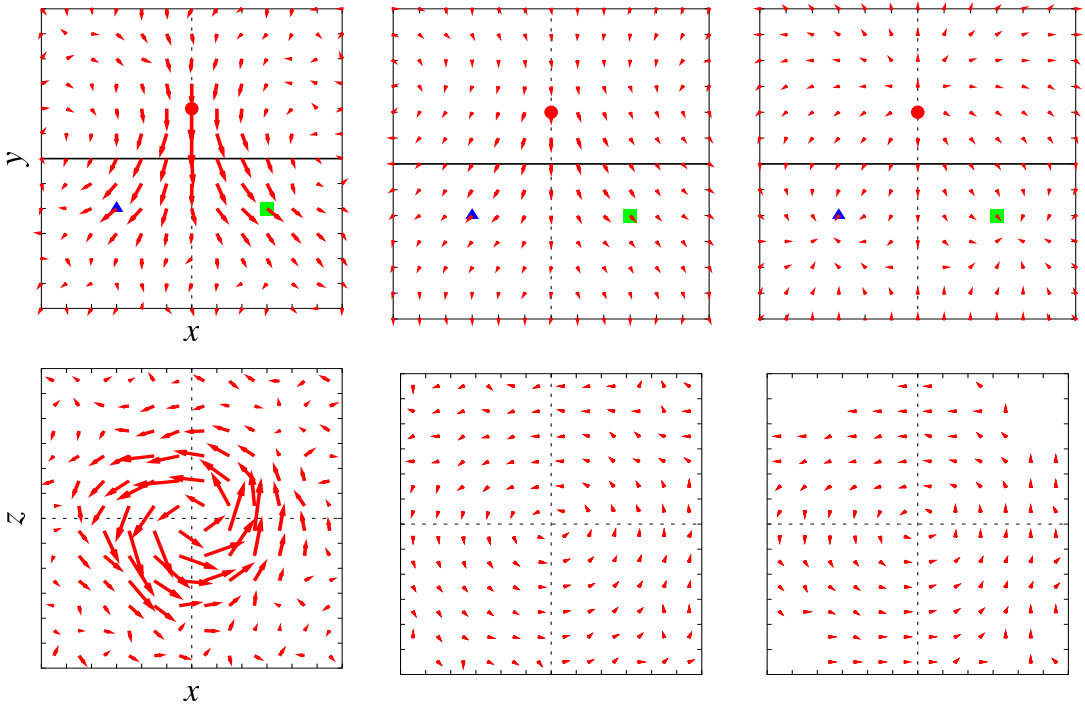}
\vspace*{0.25cm}
\caption{The color electric field (top) and monopole currents (bottom) on the
$16^3\, 8$ lattice at $\kappa=0.1335$ (left), $0.1344$ (middle) and $0.1360$
(right). The three quarks lie in (what we call) the $(x,y)$ plane. The
bottom figures show the monopole currents in the $(x,z)$ plane at the position
marked by the solid lines in the top figures.} \label{fig:R04mon}
\end{center}
\end{figure*}

In Fig.\ref{fig:R04mon} we show the monopole part of the
electric field, averaged over the color components, and the accompanying
monopole current for three values of  
$\kappa$, corresponding (from left to right) to the confined case, to $T
\approx T_c$ and to
the deconfined phase. In the confinement phase ($\kappa = 0.1335$) we find
the flux to be of {\sf Y}-shape, similar to the zero temperature case where
we used Wilson loop correlators. Note that the Polyakov lines do not have a 
{\sf Y}-shape junction like the Wilson loop does, which excludes the 
possibility that the flux is being induced by the color lines. Just below 
$T_c$ ($\kappa=0.1344$) we still see a {\sf Y}-shape flux, 
while in the deconfined phase ($\kappa=0.1360$) the electric field becomes
Coulomb-like.

\section{Conclusions}

We have studied the $3Q$ system in the maximally abelian gauge in full QCD at
zero and at finite temperature. Among the quantities we have looked at are
the abelian baryon potential as well as the flux distribution and the
action density.
While on the basis of the potential it is hard to decide whether the 
long-range potential is of $\Delta$- or {\sf Y}-type, the distribution of the
color electric field and the action density clearly shows a {\sf Y}-shape
geometry. As in the $Q\bar{Q}$ system, we identified the solenoidal monopole 
current to be responsible for squeezing the color electric flux into a narrow 
tube. Little difference to the quenched theory was found.
In the deconfined phase the flux tube disappears, and the color electric
field assumes a Coulomb-like form. Our results are in qualitative agreement 
with the predictions of the dual Ginzburg-Landau model~\cite{Koma}: 
the baryon flux has $Y$-shape, and the solenoidal monopole currents
are clearly observed.

\section*{Acknowledgements}

The dynamical gauge field configurations at $T=0$ have been generated on the
Hitachi SR8000 at LRZ Munich. We thank the operating staff for their
support. The dynamical gauge field configurations at $T>0$ have been generated 
on the Hitachi SR8000 at KEK Tsukuba. The  
analysis has largely been done on the COMPAQ Alpha Server ES40 at Humboldt 
University, as well as on the NEC SX5 at RCNP Osaka.
We wish to thank M.~N.~Chernodub, M.~M\"uller-Preussker, Y.~Koma and 
H.~Suganuma for useful
discussions. H.I. thanks the Humboldt University and Kanazawa University for
hospitality. V.B. is supported by JSPS. T.S. is supported by JSPS 
Grant-in-Aid for Scientific Research on Priority Areas 13135210 and
15340073. M.I.P. is  
supported by grants RFBR 02-02-17308, RFBR 01-02-17456, INTAS-00-00111,
DFG-RFBR 436 RUS 113/739/0, RFBR-DFG 03-02-04016 and
CRDF award RPI-2364-MO-02.

\end{document}